# Simulation and optimization of arsenic-implanted THz emitters

M.B. Johnston, J. Lloyd-Hughes, E. Castro-Camus, M.D. Fraser[1] and C. Jagadish[1],

Clarendon Laboratory, University of Oxford, Parks Road Oxford OX1 3PU, United Kingdom.
[1] Department of Electronic Materials Engineering, Research School of Physical Sciences and Engineering, Institute of Advanced Studies, Australian National University, Canberra ACT 0200, Australia.
e-mail: m.johnston@physics.ox.ac.uk

**Abstract**

We have used a three-dimensional pseudo-classical Monte Carlo simulation to investigate the effects of $As^+$ ion-implantation on pulsed terahertz radiation emitters. Devices based on surface-field emitters and photoconductive switches have been modelled. Two implantations of $As^+$ ions at 1.0 MeV and 2.4 MeV were found to produce a uniform distribution of vacancies over the volume of GaAs contributing to THz generation in these devices. We calculate that ion-implantation increases the THz bandwidth of the devices with the cost of decreasing the spectral intensity at lower THz frequencies.

**Introduction**

Coherent time-domain spectroscopy (TDS) is already becoming an important tool in many areas of spectroscopy and imaging. Recent applications have included: observing the onset of particle screening in GaAs [1]; measuring low-energy vibrational modes in oligomers [2], and non-destructive testing of space-shuttle foam [3]. To date the majority of studies have been conducted in the few-terahertz regime (i.e. with a accessible band from ~0.1-3 THz). However, for many applications, particularly those in chemistry, spectroscopy at frequencies above 3 THz would be extremely beneficial. Thus there is considerable interest in developing broadband sources and detectors of THz radiation.

Devices used to generate THz radiation for time domain spectroscopy applications typically rely on sub-picosecond laser pulses, and hence THz technology has closely followed developments in femtosecond solid-state laser technology. The two most commonly used techniques for producing THz pulses from ~100fs near-infrared laser pulses are (1) optical rectification in a non-linear crystal such as ZnTe and (2) emission from hot photo-generated carriers in the surface region of a semiconductor. Here we concentrate on the second technique, which include surface-field emitters and photoconductive switches.

Surface field emitters are simply pieces of bulk semiconductor onto which above-bandgap femtosecond pulses are incident. THz radiation results from the ultrafast separation of charge carriers by a combination of two mechanisms: screening of the semiconductor's intrinsic surface electric field, or via the photo-Dember effect [4]. These devices can be used to produce well-collimated THz beams[5].

In photoconductive switches devices charge separation is initiated using a large external electric field. The external field is typically applied in the plane of the semiconductor's surface using metal electrodes. In this way the dipole is in the optimal orientation for efficient coupling of THz radiation into free space.

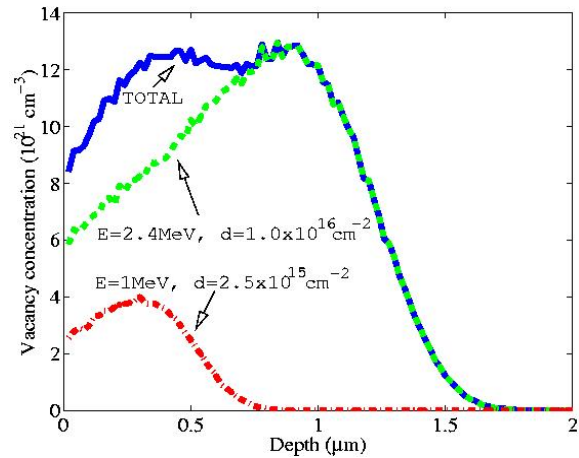

**Fig. 1**: Depth dependence of vacancies (including recoil vacancies) for a doubly ion implanted GaAs wafer. Calculated using SRIM software [8].

Photoconductive switches are some of the most efficient pulsed THz sources with electric fields of up to 9500V/m having been observed experimentally`[6]. In comparison, InAs based surface field emitters typically produce fields of ~200V/m.

A recent experimental study [7] recorded THz emission over a frequency range from 0.3THz to 20THz using low-temperature grown GaAs photoconductive emitters. Low temperature grown GaAs was found to have a larger bandwidth than similar experiments performed on semi-insulating GaAs. However LT GaAs is difficult to grow reliably. An alternative way of introducing vacancies and dislocations into semiconductors is to use ion implantation. The effects of ion implantation are well studied and precise doping and damage profiles may be created high level of accuracy and reproducibility. Furthermore, masked multi-energy ion-implantation offers the possibility of controlling the vacancy concentration in all three dimensions.

**Carrier-dynamics modelling**

We have previously developed a three dimensional pseudo classical Monte Carlo simulation for investing hot carrier dynamics in the vicinity of a semiconductor's surface [4]. The model simulates the trajectories of extrinsic and photo-generated carriers. Scattering rates and angles are calculated microscopically for each electron and hole in the simulation at each time-step in the simulation. Polar and non-polar optical phonon scattering, impurity scattering, vacancy scattering and carrier-carrier scattering mechanisms are included. The simulation has been used to establish the mechanism of magnetic-field enhanced THz generation and the surface field THz generation [4].

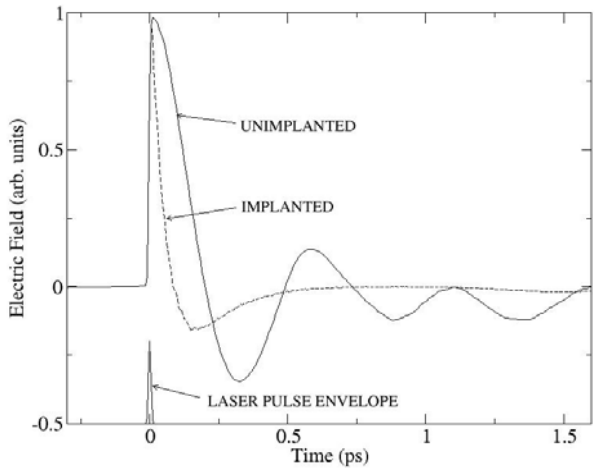

**Fig. 2:** Simulated THz electric field transient from a photoconductive switch on semi-insulating GaAs (solid line) and ion implanted GaAs, where As-vacancy scattering is included. The envelope of the exciting near-IR pulse is shown (offset) for comparison.

We simulate the effect of the ion-implantation by the introduction of a microscopic vacancy scattering mechanism, whereby carriers are scattered elastically from a spherical square well. We also include the effects of carrier trapping at antisite donor defects $As_{Ga}$ by including a characteristic capture time.

In this study we consider THz devices fabricated from a GaAs wafer implanted with a $2.5 \times 10^{15}$ cm$^{-2}$ dose of 1.0 MeV As$^+$ ions and a $1.0 \times 10^{16}$ cm$^{-2}$ dose of 2.4 MeV As$^+$ ions. Fig. 1 shows the depth dependence of vacancies, including recoil vacancies for such a wafer calculated using SRIM software [8]. It can be seen that the vacancy concentration is approximately constant over the region of the sample where 800 nm wavelength photons are absorbed. Therefore, in our simulations we may assume that the vacancy concentration is uniform throughout the simulation volume. We also assume a carrier-trapping lifetime of 0.1 ps. Details of other simulation parameters may be found in Ref. [4].

**Results**

Fig. 2 shows the results of a simulation performed on a GaAs photoconductive switch with an average applied field of 50kV/cm between its surface electrodes. These results represent the electric field transient emitted from the device (in a back-relection geometry of Ref [7]) as a function of time after photo-excitation by a 10 fs duration laser pulse (centre wavelength 800nm). The solid trace shows the emitted electric field for a device fabricated on a standard semi-insulating GaAs substrate, whereas the dashed trace simulates the device after ion-implantation and thermal annealing.

Note that the initial peak in the transient is much narrower for the ion implanted transient and that the subsequent plasmon oscillation are over-damped as a result of both vacancy scattering and carrier capture at $As_{Ga}$ defects. To consider the implications of these results for spectroscopic applications, we have calculated the power spectra of these results. The spectra in Fig 3 show that the spectral intensity of the implanted device is improved for frequencies above 6 THz. However, this significant increase in bandwidth is offset by an order of magnitude reduction in the peak spectral intensity at ~2THz.

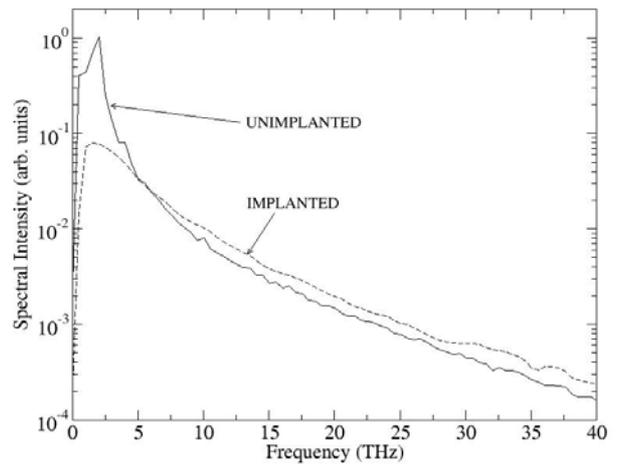

**Fig. 3:** Power spectra of the simulated data presented in Fig 2. Note that the increase in bandwidth in the implanted case, accounted for by a reduction in the peak spectral intensity. Note we do not include coherent photon oscillations in our simulation, so phonon resonances are not seen in the simulation results.

Simulations were also performed on surface-field THz emitters. These results showed a similar trend to the photoconductive switch results with the spectral bandwidth of the device being increased as a result of implantation. However a degradation of the peak intensity was also observed in these devices.

We expect our model to also be suitable for simulating THz radiation emission from low-temperature grown GaAs emitters, since arsenics antisite defect are also the dominant defect in these materials. Interestingly, the power spectra of our modelled ion implanted photoconductive switches are consistent with the experimentally measured spectra given in Ref [7] for low-temperature grown GaAs emitters.